\documentstyle{article}
\begin{document}
\title{
Scattering Matrix of the SU(n) Gauge Theory with \\ Explicit Gauge Mass Term}
\author{Yang Ze-sen, Li Xianhui, Zhou Zhining and Zhong Yushu \\
Department of Physics,  Peking University, Beijing 100871, CHINA }
\date{\today}
\maketitle
\begin{abstract}
   Based on the renormalisability of the SU(n) theory with massive gauge
bosons, we start with the path integral of the generating
functional for the renormalized Green functions and develop a
method to construct the scattering matrix so that the unitarity is
evident. By using as basical variables the renormalized field
functions and defining the unperturbed Hamiltonian operator $H_0$
that, under the Lorentz condition, describes the free particles of
the initial and final states in scattering processes, we form an
operator description with which the renormalized Green functions
can be expressed as the vacuum expectations of the time ordered
products of the Heisenberg operators of the renormalized field
functions, that satisfy the usual equal time commutation or
anticommutation rules. From such an operator description we find a
total Hamiltonian $\widetilde{H}$ that determine the time
evolution of the Heisenberg operators of the renormalized field
functions. The scattering matrix is nothing but the matrix of the
operator $U(\infty, -\infty)$, which describes the time evolution
from
 $-\infty$ to $\infty$ in the interaction picture specified by
 $\widetilde{H}$ and $H_0$, respect to a base formed by the physical eigen
states of $H_0$. We also explain
the asymptotic field viewpoint of constructing the scattering matrix within
our operator description. Moreover, we find a formular to express the
scattering matrix elements in terms of the truncated renormalized Green
functions.
\end{abstract}
\begin{center}
PACS numbers: 03.65.Db, 03.80.+r, 11.20.Dj
\end{center}
\vspace{3mm}
\newpage
\begin{center}
{\bf I}.\ \ Introduction
\end{center} \par
    Although the SU(n) theory with a mass term of the  gauge fields was
considered to be in general nonrenormalisable for a long time, it was
repeatedly studied (see for example Refs. [1-8]). In particular, we have
proven the renormalisability of the theory under the original expression
of the mass term of the  gauge fields [8]. Therefore it becomes possible to
to deeply studied the scattering matrix of such theories. In the present
paper, instead of checking the negative arguments concerning the unitarity
we will develop a method to construct the scattering matrix with the
renormalized field functions as basical variables so that the unitarity is
evident.  \par
     Owing to the renormalisability of theory, one can start with the
path integral of the generating functional for the renormalized
Green functions and form  an operator description with the
renormalized field functions  as basical variables. This  means
that the Heisenberg operators of the renormalized field functions
satisfy the usual equal time commutation or anticommutation rules
and the renormalized Green functions can be expressed as the
vacuum expectations of the time ordered products of such
Heisenberg field operators. On the other hand starting from the
path integral of the generating functional for the unrenormalized
Green functions, one can form another operator description with
the bare field functions as basical variables. It should be
emphasized that such two operator descriptions are of different
types. In particular, the field operators belonging to the two
operator descriptions operate on different state spaces and can
not be transformed into each other by unitary transformations or
by scale transformations.
\par
     According to the very idea of renormalisation we decide to choose the
first kind of operator description.  With the renormalized field functions
and the renormalized parameters we will define a non-interacting Lagrangian
${\cal L}_0$ which, under the Lorentz condition, describes the free particles
of the initial and final states in scattering processes. Next we will find an
unperturbed Hamiltonian $H_0$ operation in a way used in the Gupta-Bleuler
quantization and define a Quasi-Gupta-Bleuler subspace ($QGB$) which is
spanned by all the physical eigen states of $H_0$.
By giving an operator expression for the generating functional for the Green
functions of the free fields, we will able to do the same for the generating
functional for the renormalized full Green functions. Then we can express
the renormalized Green functions as the vacuum expectations of the time
odered products and find the total Hamiltonian $\widetilde{H}$ that determine the time evolution of
the Heisenberg operators of the renormalized field functions. With the $QGB$
subspace and the operator $U(\infty,-\infty)$, which describes the time
evolution from  $-\infty$ to $\infty$ in the interaction picture specified
by $\widetilde{H}$ and $H_0$, we can easily  define  the scattering matrix
and make the unitarity evident.
\par
   The operator description will be given in section $2$. In section $3$
we will explain the definition and the unitarity of the scattering matrix
and briefly describe the asymptotic field viewpoint based on our operator
description. In section $4$ a formular will be found to express the
scattering matrix elements in terms of the truncated renormalized Green
functions. Concluding remarks will be given in the final section.
\par
\vspace{5mm}
\setcounter{equation}{0}
\def\theequation{2.\arabic{equation}}
\begin{center}
{\bf II}.\ \ Operator description with renormalized field functions
             as basical variables
\end{center} \par
    Assume that the theory has been renormalized under a proper
renormalization scheme so that the renormalized parameters such as the masses
represent the physical ones. The generating functional for the renormalized
Green functions can be expressed as
\begin{eqnarray}
{\cal Z}^{[R]}[J,\overline{\chi},\chi,\overline{\eta},\eta]
= \frac{1}{N}
\int {\cal D}[{\cal A},\overline{\omega},\omega,\overline{\psi},\psi]
\prod_{a',x'} \delta \left(\partial^{\mu}A_{a' \mu}(x') \right)
 {\rm exp} \Big\{ {\rm i} \big( \overline{I}_{{\rm eff}} + I_s \big) \Big\}\,,
\end{eqnarray}
where $A_{a \mu},\omega_a,\overline{\omega}_a$ and
$\psi,\overline{\psi}$ stand for the renormalized gauge fields,
ghost fields and material fields, all the parameters are also
renormalized ones, $N$ is a constant to make
 ${\cal Z}^{[R]}[0,0,0,0,0]$ equal to $1$,
 $I_s$ is the source term
\begin{eqnarray*}
&&I_s = \int d^4x \big[ J_a^{\mu}(x) A_{a \mu}(x)
   + \overline{\chi}_a(x)\omega_a(x) + \overline{\omega}_a(x) \chi_a(x) \\
&& \ \ \ \ \ \ \
+ \overline{\eta}_a(x) \psi_a(x) + \overline{\psi}_a(x) \eta_a(x) \big]\,.
\end{eqnarray*}
The action $\overline{I}_{{\rm eff}}$ contains the counterterm and is
formed by the following Lagrangian
\begin{eqnarray}
&&\overline{\cal L}_{{\rm eff}} = -\frac{1}{4} F_{a \mu \nu} F_a^{\mu \nu}
         + \frac{1}{2} M^2 A_{a \mu}A_{a}^{\mu}
           + {\cal L}_{\psi} + {\cal L}_{\psi A} \nonumber\\
&&\ \ \ \ \ \ \ \
 + \big(- \partial_{\mu}\overline{\omega}_a(x)\big) D^{\mu}_{ab} \omega_b(x)
    + {\cal L}_{count} \,.
\end{eqnarray}
    The free evolution of the physical particles of the initial and final
states in scattering processes should be described by combining
the Lorentz condition with the Lagrangian
\begin{eqnarray}
{\cal L}_0(x) = -\frac{1}{4} F^{[0]}_{a \mu \nu} F_a^{[0]\mu \nu}
+ \frac{1}{2} M^2 A_{a \mu}(x) A_{a}^{\mu}(x) + {\cal L}_{\psi}(x)
+ \big(- \partial_{\mu}\overline{\omega}_a(x)\big) D^{\mu}_{ab} \omega_b(x)\,,
\end{eqnarray}
where
$$
 F^{[0]}_{a \mu \nu} = \partial_{\mu}A_{a \nu}-\partial_{\nu}A_{a \mu}\,.
$$
In order to describe such free particles by an operator method, we replace
(2.3) by
\begin{eqnarray}
\widetilde{{\cal L}}_0(x) = {\cal L}_{GB}(x)
+ \frac{1}{2} M^2 A_{a \mu}(x) A_{a}^{\mu}(x) + {\cal L}_{\psi}(x)
+ \big(- \partial_{\mu}\overline{\omega}_a(x)\big) D^{\mu}_{ab} \omega_b(x)\,,
\end{eqnarray}
where ${\cal L}_{GB}(x)$ is a Gupta--Bleuler Lagrangian
$$
{\cal L}_{GB}(x)
       =  -\frac{1}{4} F^{[0]}_{a \mu \nu} F_a^{[0]\mu \nu}
         - \frac{1}{2} \big( \partial^{\mu} A_{a \mu} \big)^2 \,.
$$
Then we perform a canonical quantization with $\widetilde{{\cal
L}}_0(x)$ and define the corresponding Hamiltonian operator $H_0$
and field operators. Moreover we have to choose the so-called
physical states and define a physical initial-final state subspace
according to Lorentz condition and the residual symmetry. This
subspace is spanned by all the physical eigen states of $H_0$ and
can be formed in a similar method used in the Gupta--Bleuler
quantization although there are differences caused by the gauge
field mass term (see the Appendix or Ref. [9]). We will simply
call such a subspace as Quasi-Gupta-Bleuler or $QGB$ subspace. If
we can find a total Hamiltonian which determines the time
evolution of the Heisenberg operators of the renormalized field
functions, we will be able to construct the scattering matrix with
the help of the $QGB$ subspace and the time evolution operators
$U(\infty,-\infty)$ which describes the time evolution from
$-\infty$ to $+\infty$ in the interaction picture defined by the
total Hamiltonian and $H_0$. It will be seen that these tentative
ideas are correct. However, we will have to introduce the
$\xi-$gauge fixing term and construct an appropriate operator
expression for the generating functional for the renormalized
Green functions.  \par
    With the help of the the gauge fixing term
${\cal L}_{GF} = -\frac{1}{2 \xi}\big( \partial^{\mu} A_{a \mu} \big)^2 $
one can express (2.1) as
\begin{eqnarray}
{\cal Z}^{[R]}[J,\overline{\chi},\chi,\overline{\eta},\eta]
 = \lim_{\xi \rightarrow 0}
    {\cal Z}^{[R]}_{\xi}[J,\overline{\chi},\chi,\overline{\eta},\eta] \,,
\end{eqnarray}
where
\begin{eqnarray}
{\cal Z}^{[R]}_{\xi}[J,\overline{\chi},\chi,\overline{\eta},\eta]
= \frac{1}{N_{\xi}}
\int {\cal D}[{\cal A},\overline{\omega},\omega,\overline{\psi},\psi]
 {\rm exp} \Big\{ {\rm i} \big( I_{{\rm eff}} + I_s \big) \Big\}\,,
\end{eqnarray}
where $ N_{\xi}$ is a constant to make ${\cal Z}^{[R]}_{\xi}[0,0,0,0,0]$
equal to $1$, and
\begin{eqnarray}
 I_{{\rm eff}} = \overline{I}_{{\rm eff}}
                 + \int d^4x {\cal L}_{GF}(x)  \,.
\end{eqnarray}
Needles to say, quantities defined by using
${\cal Z}^{[R]}_{\xi}[J,\overline{\chi},\chi,\overline{\eta},\eta] $ depend
on $\xi$, but this parameter will often be omitted.
\par
   Let $\widetilde{{\cal Z}}^0[J,\overline{\chi},\chi,\overline{\eta},\eta]$
be the generating functional for the Green functions defined by the effective
Lagrangian $\widetilde{{\cal L}}_0$, namely
$$
\widetilde{{\cal Z}}^0[J,\overline{\chi},\chi,\overline{\eta},\eta]
= \frac{1}{N_0}
  \int {\cal D}[{\cal A},\overline{\omega},\omega,\overline{\psi},\psi]
  \prod_{a',x'} \delta \left(\partial^{\mu}A_{a' \mu}(x') \right)
  {\rm exp} \Big\{ {\rm i} \big( \int d^4x \widetilde{{\cal L}}_0(x)
        + I_s \big) \Big\}\,,
$$
where $N_0$ is a constant to make $\widetilde{{\cal
Z}}_0[0,0,0,0,0]$ equal to $1$.  One thus has
\begin{eqnarray}
\widetilde{{\cal Z}}^0[J,\overline{\chi},\chi,\overline{\eta},\eta]
    &=& \langle 0| {\bf T}{\rm exp}\Big\{ {\rm i}
       \int d^4x \big[ J_a^{\mu}(x) A_{a \mu}(x) \nonumber\\
       &&+\, \overline{\chi}_a(x)\omega_a(x)
        + \overline{\omega}_a(x) \chi_a(x)
        + \overline{\eta}_a(x)\psi_a(x) + \overline{\psi}_a(x)\eta_a(x)
          \big]\Big\} |0\rangle\,,
\end{eqnarray}
where $A_{a \mu}(x) $,$\omega_a(x)$,$\psi_a(x)$,$\cdots$ are field
operators satisfying the usual equal time commutation or
anticommutation rules and their time evolution are determined by
$H_0$. The state $|0\rangle$ is the eigen state of $H_0$ belonging
to the lowest eigenvalue (assigned to be zero). Define
$\widetilde{{\cal H}}_I$ so that
\begin{eqnarray}
I_{{\rm eff}}
      = - \int d^4x \widetilde{{\cal H}}_I(x)
        + \int d^4x \widetilde{{\cal L}}_0(x) \,.
\end{eqnarray}
Then one gets from (2.6)
\begin{eqnarray}
&&{\cal Z}^{[R]}_{\xi} [J,\overline{\chi},\chi,\overline{\eta},\eta]
   \nonumber\\
&& \ \ \ \ \
  \propto  {\rm exp} \left\{
       -{\rm i} \int d^4x \widetilde{{\cal H}}_I
\left(
      \frac{\delta}{{\rm i} \delta J(x)},
      \frac{\delta}{{\rm i} \delta \overline{\chi}(x)},
      \frac{\delta}{(-{\rm i}) \delta \chi(x)},
      \frac{\delta}{{\rm i} \delta \overline{\eta}(x)},
      \frac{\delta}{(-{\rm i}) \delta \eta(x)}
\right)
\right\}
\widetilde{{\cal Z}}_0[J,\overline{\chi},\chi,\overline{\eta},\eta]
  \nonumber\\
&&\ \ \ \ \ \propto
\langle\,0 |{\bf T}\,\Big( {\rm exp}\Big\{{\rm i} \int d^4x
\big[ J_a^{\mu}(x) A_{a \mu}(x)
        + \overline{\chi}_a(x)\omega_a(x) \nonumber\\
&&\ \ \ \ \ \ + \overline{\omega}_a(x) \chi_a(x)
        + \overline{\eta}_a(x)\psi_a(x) + \overline{\psi}_a(x)\eta_a(x)
 \big] \Big\}
{\bf T}\,{\rm exp}\Big\{-{\rm i}\,\int d^4x\,\widetilde{{\cal H}}_I(x)\Big\}
\Big) | 0\,\rangle \,.
\end{eqnarray}
where
 $\widetilde{{\cal H}}_I\left(
      \frac{\delta}{{\rm i} \delta J(x)},
      \frac{\delta}{{\rm i} \delta \overline{\chi}(x)},
      \frac{\delta}{-{\rm i} \delta \chi(x)},
      \frac{\delta}{{\rm i} \delta \overline{\eta}(x)},
      \frac{\delta}{-{\rm i} \delta \eta(x)}
 \right)$
is a differential operator  obtained from  $\widetilde{{\cal H}}_I(x)$
 by replacing the field functions
 $ A_{a \mu}(x)$,
 $\omega_a(x)$,
 $\overline{\omega}_a(x)$,
 $\psi_{a \alpha}(x)$ and
 $\overline{\psi}_{a \beta}(x)$
with
 $ \frac{\delta}{{\rm i} \delta J_a^{\mu}(x)}$,
 $ \frac{\delta}{{\rm i} \delta \overline{\chi}_a(x)}$,
 $ \frac{\delta}{-{\rm i} \delta \chi_a(x)}$,
 $ \frac{\delta}{{\rm i} \delta \overline{\eta}_{a \alpha}(x)}$ and
 $ \frac{\delta}{-{\rm i} \delta \eta_{a \beta}(x)} $
respectively. The operator $\widetilde{{\cal H}}_I$ appearing in
the last line is obtained from the classical quantity by replacing
the field functions with operators
 $ A_{a \mu}(x)$, $\omega_a(x)$,
 $\overline{\omega}_a(x)$,
 $\psi_{a \alpha}(x)$ and
 $\overline{\psi}_{a \beta}(x)$ and then taking the normal products with
respect to $|0\rangle$. Each of the two proportion coefficients in
(4.10) should make ${\cal Z}^{[R]}_{\xi}[0,0,0,0,0]$ equal to $1$.
It is evident that the time evolution of $\widetilde{{\cal H}}_I$
is determined by $H_0$:
\begin{eqnarray}
 \widetilde{{\cal H}}_I(t)
   =  e^{{\rm i}H_0t} \widetilde{{\cal H}}_I(0) e^{-{\rm i} H_0t} \,.
\end{eqnarray}
Therefore, the operator
$$  {\bf T}\,{\rm exp}\Big\{
        -{\rm i}\,\int d^4x\,\widetilde{{\cal H}}_I(x)\Big\} $$
is the time evolution operator $U(\infty,-\infty)$ in the
interaction picture defined by $H_0$ and the interaction $\int
d^3x \widetilde{{\cal H}}_I(0)$, namely
\begin{eqnarray}
 {\bf T}\,{\rm exp}\Big\{-{\rm i} \int d^4x\,\widetilde{{\cal H}}_I(x)\Big\}
  = U(\infty,-\infty) \,,
\end{eqnarray}
and
\begin{eqnarray}
&& U(t,t_0) =  e^{{\rm i}H_0t} e^{-{\rm i}
              \widetilde{H}(t-t_0)} e^{-{\rm i} H_0 t_0} \,,
\\
&& \widetilde{H} = H_0 + \int d^3x \widetilde{{\cal H}}_I(0) \,.
\end{eqnarray}
It follows that (2.10) can be written as
\begin{eqnarray}
{\cal Z}^{[R]}_{\xi} [J,\overline{\chi},\chi,\overline{\eta},\eta]
 &\propto & \langle\,0\,|U(\infty,0)
\Big({\bf T}\,{\rm exp}\Big\{ {\rm i}
\int d^4x \Big[
        J_a^{\mu}(x) A^h_{a \mu}(x)
       + \overline{\chi}_a(x)\omega^h_a(x) \nonumber\\
       &&+\, \overline{\omega}^h_a(x) \chi_a(x)
        + \overline{\eta}_a(x)\psi^h_a(x) + \overline{\psi}^h_a(x)\eta_a(x)
\Big]\Big\}\Big)
U(0,-\infty) |\,0\,\rangle \,,
\end{eqnarray}
The proportion coefficient should make
 ${\cal Z}^{[R]}_{\xi}[0,0,0,0,0]$ equal to $1$.
The time evolution of the operators $A^h_{a \mu}, \cdots$ are determined by
 $\widetilde{H} $:
\begin{eqnarray*}
&& A^h_{a \mu}(t)
 =  e^{{\rm i} \widetilde{H}t} A_{a \mu}(0) e^{-{\rm i} \widetilde{H}t} \,,
\\
&& \omega^h_a(t)
 = e^{{\rm i} \widetilde{H}t} \omega_a(0) e^{-{\rm i} \widetilde{H}t} \,,
\\
&& \overline{\omega}^h_a(t)
 = e^{{\rm i}\widetilde{H}t}\overline{\omega}_a(0)
                             e^{-{\rm i}\widetilde{H}t} \,,
\\
&& \psi^h_a (t)
 = e^{{\rm i}\widetilde{H}t} \psi_a(0) e^{-{\rm i} \widetilde{H}t} \,,
\\
&& \overline{\psi}^h_a(t)
 = e^{{\rm i}\widetilde{H}t}\overline{\psi}_a(0)e^{-{\rm i}\widetilde{H}t} \,.
\end{eqnarray*}
Therefore $\widetilde{H}$ is the total Hamiltonian operator and $
A^h_{a \mu}(x)$, $\omega^h_a(x)$, $ \overline{\omega}^h_a(x)$, $
\psi^h_a(x)$ and $ \overline{\psi}^h_a(x)$ stand for the
Heisenberg operators of the renormalized field functions. Thus
(2.15) means that the renormalized Green functions can be
expressed as the vacuum expectations of the time ordered products
of the Heisenberg operators of the renormalized field functions.
It should be emphasized that these field operators satisfy the
usual equal time commutation or anticommutation rules.
\par
\vspace{5mm}
\setcounter{equation}{0}
\def\theequation{3.\arabic{equation}}
\begin{center}
{\bf III}.\ \ The scattering matrix
\end{center} \par
    One can choose to specify the initial condition at $t=-\infty$ or at
$t=+\infty$ for state vectors in the interaction picture and constructing
the scattering theory. Therefore  for an arbitrary state $|\Psi_{GB}\rangle$
in the $QGB$ subspace there is a state $|\Psi'_{GB}\rangle$
in this subspace so that
\begin{eqnarray}
 U(t,\infty) | \Psi'_{GB} \rangle = U(t,-\infty)| \Psi_{GB} \rangle \,.
\end{eqnarray}
Conversely, for an arbitrary state $|\Psi'_{GB}\rangle \in QGB$, there is
a state $|\Psi_{GB}\rangle \in QGB$ that satisfies this equation. Next,
states $U(t,\pm\infty) |\Psi_{GB} \rangle $  are normalized just
like $|\Psi_{GB} \rangle $  one thus has
\begin{eqnarray}
 U(\pm\infty,t) U(t,\pm\infty) | \Psi_{GB} \rangle = | \Psi_{GB} \rangle \,.
\end{eqnarray}
This and (3.1) also indicate that the operators $U(\pm\infty,\mp\infty)$
conserve invariant the $QGB$ subspace. We therefore express the elements of
the scattering matrix as
\begin{eqnarray}
&& S_{fi} = \lim_{\xi \rightarrow 0}
           \frac{ \langle f | U(\infty,-\infty) | i \rangle }
                { \langle 0 | U(\infty,-\infty) | 0 \rangle } \,,
\end{eqnarray}
where  $| i \rangle $ and $| f \rangle $ are within the QGB subspace and
are the eigen states of $H_0$.
\par
    From (3.1) and (3.2), for an arbitrary state $|\Psi_{GB}\rangle \in QGB$
one has
\begin{eqnarray*}
&&  U(0,-\infty)|\Psi_{GB} \rangle = U(0,\infty)|\Psi'_{GB} \rangle \,,\\
&& U(\infty,-\infty) |\Psi_{GB} \rangle = |\Psi'_{GB} \rangle \,,\\
&& U(-\infty,\infty) |\Psi'_{GB} \rangle =  |\Psi_{GB} \rangle\,.
\end{eqnarray*}
Thus,
$$
 U(-\infty,\infty) U(\infty,-\infty)|\Psi_{GB} \rangle
     =  U(-\infty,\infty) |\Psi'_{GB} \rangle
     =  |\Psi_{GB} \rangle\,.
$$
Similarly for arbitrary $|\Psi'_{GB}\rangle \in QGB$,
$$
 U(\infty,-\infty) U(-\infty,\infty)|\Psi'_{GB} \rangle
     =  |\Psi'_{GB} \rangle\,.
$$
These ensure the unitarity of the scattering matrix.
\par
      With the help of the time evolution operator in the interaction
picture we can also define the operators of so-called asymptotic fields
which will be denoted by
 $A_{a \mu,{\rm in}}(x)$, $ A_{a \mu,{\rm out}}(x)$, $\psi_{a,{\rm in}}(x)$
 and $ \psi_{a,{\rm out}}(x)$, $\cdots$. These are free fields, and when
$-t$ is very large each in-operator proportional to the Heisenberg
operator of the renormalized field function. Actually each
proportion coefficient should be equal to $1$ in order not to
destroy the usual equal time commutation or anticommutation rules.
On the other hand when $t$ is very large each out-operator is
equal to the Heisenberg operator of the renormalized field
function.  Using
\begin{eqnarray*}
&& A^h_{a \mu}(x) = U(0,t) A_{a \mu}(x) U(t,0) \,,
\\
&& \psi^h_a(x) = U(0,t) \psi_a(x) U(t,0) \,,
\\
&& \omega^h_a(x) = U(0,t) \omega_a(x) U(t,0)  \,,
\end{eqnarray*}
and noticing that when $|t|$ is large enough, $U(0,t)$ and $U(t,0)$ are
independent of $t$, one gets
\begin{eqnarray}
&& A_{a \mu,{\rm in}}(x) = U(0,-\infty) A_{a \mu}(x) U(-\infty,0) \,,
\\
&& \psi_{a,{\rm in}}(x) = U(0,-\infty) \psi_a(x) U(-\infty,0) \,,
\\
&& A_{a \mu,{\rm out}}(x) = U(0,\infty) A_{a \mu}(x) U(\infty,0) \,,
\\
&& \psi_{a,{\rm out}}(x) = U(0,\infty) \psi_a(x) U(\infty,0) \,.
\end{eqnarray}
Since the asymptotic fields are free ones, these are valid for any $t$.
The formulae for the other fields are similar.
\par
    If the in--states and out--states are defined by
\begin{eqnarray*}
&& |0,{\rm in} \, \rangle = \frac{ U(0,-\infty) |0 \rangle}
               {\sqrt{\langle\, 0| U(\infty,-\infty) |0 \,\rangle}} \,,
\\
&& |0,{\rm out} \,\rangle = \frac{  U(0,\infty) |0 \rangle }
               {\sqrt{\langle\, 0| U(\infty,-\infty) |0 \,\rangle}}\,,
\\
&& |i,{\rm in} \, \rangle = \frac{ U(0,-\infty) |i \rangle }
               {\sqrt{\langle\, 0| U(\infty,-\infty) |0 \,\rangle}}\,,
\\
&& |i,{\rm out} \,\rangle = \frac{ U(0,\infty) |i \rangle }
               {\sqrt{\langle\, 0| U(\infty,-\infty) |0 \,\rangle}}\,,
\end{eqnarray*}
then (3.9) can be written as
\begin{eqnarray}
&& S_{fi} = \lim_{\xi \rightarrow 0} \langle f,{\rm out}| i,{\rm in}\rangle
  = \lim_{\xi \rightarrow 0}
    \langle\, f,{\rm in}| U(0,-\infty) U(\infty,0)|i,{\rm in} \,\rangle \,.
\end{eqnarray}
\par
\vspace{5mm}
\setcounter{equation}{0}
\def\theequation{4.\arabic{equation}}
\begin{center}
{\bf IV}.\ \ An expression for the scattering matrix elements in terms \\
              of the truncated renormalized Green functions
\end{center}
\par
    In the following we will derive a formula to express the elements of the
scattering matrix in terms of the truncated renormalized Green functions.
Define the following operator
\begin{eqnarray}
U(J,\overline{\chi},\chi,\overline{\eta},\eta) &=&
{\bf T}\,\Big(U(\infty,-\infty){\rm exp}\Big\{{\rm i} \int d^4x
\big[ J_a^{\mu}(x) A_{a \mu}(x)
        + \overline{\chi}_a(x)\omega_a(x) \nonumber\\
  &&\ \ \ \ \ \ \ \ + \overline{\omega}_a(x) \chi_a(x)
        + \overline{\eta}_a(x)\psi_a(x) + \overline{\psi}_a(x)\eta_a(x)
 \big] \Big\} \Big) \,.
\end{eqnarray}
We thus have
$$
 U(0,0,0,0,0) = U(\infty,-\infty)\,,
$$
and
\begin{eqnarray}
& \langle 0|
 U(J,\overline{\chi},\chi,\overline{\eta},\eta)|0 \rangle  /
     \big(\langle 0|U(\infty,-\infty) |0 \rangle \big)
 = {\cal Z}^{[R]}_{\xi}[J,\overline{\chi},\chi,\overline{\eta},\eta]\,.
\end{eqnarray}
Analogous to (2.10), we also have
\begin{eqnarray}
&&U(J,\overline{\chi},\chi,\overline{\eta},\eta) \nonumber\\
 &&\ \ =\, {\rm exp} \left\{
       -{\rm i} \int d^4x \widetilde{{\cal H}}_I
\left(
      \frac{\delta}{{\rm i} \delta J(x)},
      \frac{\delta}{{\rm i} \delta \overline{\chi}(x)},
      \frac{\delta}{(-{\rm i}) \delta \chi(x)},
      \frac{\delta}{{\rm i} \delta \overline{\eta}(x)},
      \frac{\delta}{(-{\rm i}) \delta \eta(x)}
\right)
\right\}
\widetilde{U}^{(0)}(J,\overline{\chi},\chi,\overline{\eta},\eta)
\,,\ \ \ \
\end{eqnarray}
where
\begin{eqnarray*}
\widetilde{U}^{(0)}(J,\overline{\chi},\chi,\overline{\eta},\eta)
&=&{\bf T}\, {\rm exp}\Big\{{\rm i} \int d^4x
\big[ J_a^{\mu}(x) A_{a \mu}(x)
        + \overline{\chi}_a(x)\omega_a(x) \\
   &&\ \ \ \ \ + \overline{\omega}_a(x) \chi_a(x)
        + \overline{\eta}_a(x)\psi_a(x) + \overline{\psi}_a(x)\eta_a(x)
 \big] \Big\} \,.
\end{eqnarray*}
Define ${\cal H}_I(x)$ and ${\cal L}'_{GF}(x)$ so that
\begin{eqnarray*}
&I_{{\rm eff}}
      = - \int d^4x {\cal H}_I(x)
        + \int d^4x \big\{{\cal L}_{GF}(x) + {\cal L}_0(x)\big\} \,, \\
&{\cal L}'_{GF}
= {\cal L}_{GF} + \frac{1}{2} \big( \partial^{\mu} A_{a \mu} \big)^2
=\big(\frac{1}{2}-\frac{1}{2\xi}\big)\big(\partial^{\mu} A_{a\mu} \big)^2 \,.
\end{eqnarray*}
We can write (4.3) as
\begin{eqnarray}
&&U(J,\overline{\chi},\chi,\overline{\eta},\eta) \nonumber\\
 &&\ \ \ =\, {\rm exp} \left\{
       -{\rm i} \int d^4x {\cal H}_I \!
\left(
      \frac{\delta}{{\rm i} \delta J(x)},
      \frac{\delta}{{\rm i} \delta \overline{\chi}(x)},
      \frac{\delta}{(-{\rm i}) \delta \chi(x)},
      \frac{\delta}{{\rm i} \delta \overline{\eta}(x)},
      \frac{\delta}{(-{\rm i}) \delta \eta(x)}
\right)
\right\}
 U^{(0)}(J,\overline{\chi},\chi,\overline{\eta},\eta) \,,\ \ \ \ \
\end{eqnarray}
where
\begin{eqnarray}
U^{(0)}(J,\overline{\chi},\chi,\overline{\eta},\eta)
 =\, {\rm exp} \left\{
       {\rm i}\!\int d^4x {\cal L}'_{GF}\!
\left(\frac{\delta}{{\rm i} \delta J(x)} \right) \right\}
\widetilde{U}^{(0)}(J,\overline{\chi},\chi,\overline{\eta},\eta) \,.
\end{eqnarray}
\par
Noticed that
$\langle 0| U^{(0)}(J,\overline{\chi},\chi,\overline{\eta},\eta)|0\rangle $
is proportional to the generating functional for the Green functions defined
by the effective Lagrangian ${\cal L}_0 + {\cal L}_{GF}$ and the latter
can be explicitly expressed as
\begin{eqnarray*}
{\cal Z}^{(0)}_{\xi}[J,\overline{\chi},\chi,\overline{\eta},\eta]
&=& {\rm exp}\Big\{ -\frac{{\rm i}}{2} \int d^4x \int d^4y
 J_a^{\mu}(x) D_{\mu \nu}^{ab}(x-y) J_b^{\nu}(y) \Big\} \\
 &&\ \ \ \times {\rm exp}\Big\{ -{\rm i} \int d^4x \int d^4y
      \overline{\chi}_a(x) C_{ab}(x-y) \chi_b(y) \Big\} \\
 &&\ \ \ \times {\rm exp}\Big\{ -{\rm i} \int d^4x \int d^4y
      \overline{\eta}_a(x) S_{ab}(x-y) \eta_b(y) \Big\} \,,
\end{eqnarray*}
where i$D_{\mu \nu}^{ab}(x-y)$, i$C_{ab}(x-y)$ and i$S_{ab}(x-y)$ are
the propagators. Particularly, one has
\begin{eqnarray}
&& D_{\mu \nu}^{ab}(k) = \frac{-1}
{k^2 - M^2 + {\rm i}\epsilon}
     \Big\{ g_{\mu \nu} +(\xi-1) \frac{ k_{\mu}k_{\nu}}
          {k^2 - \xi M^2 + {\rm i}\epsilon} \Big\}\delta_{ab} \,,
\\
&& [D(k)^{-1}]^{\mu \nu}_{ab}
        = \Big\{-(k^2 - M^2 + {\rm i}\epsilon) g^{\mu \nu}
          + \big(1-\frac{1}{\xi} \big) k^{\mu}k^{\nu}\Big\}\delta_{ab} \,.
\end{eqnarray}
\par
    As for the operator
 $ U^{(0)}(J,\overline{\chi},\chi,\overline{\eta},\eta)$, one has
\begin{eqnarray*}
&& U^{(0)}(J,\overline{\chi},\chi,\overline{\eta},\eta) \\
&&\ \ \ \ \propto \ : {\rm exp} \Big \{ - \int d^4x \int d^4y
      \Big( A_{a\mu}(x)\big[ D(x-y)^{-1}\big]_{ab}^{\mu \nu}
            \frac{\delta}{\delta J_b^{\nu}(y)} \\
&&\ \ \ \ \ \ \
    + \overline{\psi}_{b\beta}(x)\big[S(x-y)^{-1} \big]_{b\beta,a\alpha}
     \frac{\delta}{\delta \overline{\eta}_{a\alpha}(y)}
     + \big[S(x-y)^{-1} \big]_{b\beta,a\alpha} \psi_{a\alpha}(y)
     \frac{\delta}{\delta \eta_{b\beta}(x)} \\
&&\ \ \ \ \ \ \
    + \overline{\omega}_b(x)\big[C(x-y)^{-1} \big]_{b a}
     \frac{\delta}{\delta \overline{\chi}_a(y)}
     + \big[C(x-y)^{-1} \big]_{b a} \omega_a(y)
     \frac{\delta}{\delta \chi_b(x)}
     \Big)\Big \} :
{\cal Z}^{(0)}_{\xi}[J,\overline{\chi},\chi,\overline{\eta},\eta] \,.
\end{eqnarray*}
Substituting this in (4.4) we get
\begin{eqnarray}
&& U(J,\overline{\chi},\chi,\overline{\eta},\eta)/
     \big(\langle 0|U(\infty,-\infty) |0 \rangle \big) \nonumber\\
&&\ \ \ \ = \,: {\rm exp} \Big \{ - \int d^4x \int d^4y
      \Big( A_{a\mu}(x)\big[ D(x-y)^{-1}\big]_{ab}^{\mu \nu}
            \frac{\delta}{\delta J_b^{\nu}(y)}  \nonumber\\
&&\ \ \ \ \ \
    + \overline{\psi}_{b\beta}(x)\big[S(x-y)^{-1} \big]_{b\beta,a\alpha}
     \frac{\delta}{\delta \overline{\eta}_{a\alpha}(y)}
     + \big[S(x-y)^{-1} \big]_{b\beta,a\alpha} \psi_{a\alpha}(y)
     \frac{\delta}{\delta \eta_{b\beta}(x)} \nonumber\\
&&\ \ \ \ \ \
    + \overline{\omega}_b(x)\big[C(x-y)^{-1} \big]_{b a}
     \frac{\delta}{\delta \overline{\chi}_a(y)}
     + \big[C(x-y)^{-1} \big]_{b a} \omega_a(y)
     \frac{\delta}{\delta \chi_b(x)}
     \Big)\Big \} :
{\cal Z}^{[R]}_{\xi}[J,\overline{\chi},\chi,\overline{\eta},\eta] \,.\ \ \
\end{eqnarray}
Denoting by i$\overline{D}_{\mu \nu}^{ab}(x-y)$,
i$\overline{C}_{ab}(x-y)$ and i$\overline{S}_{ab}(x-y)$ the full
propagators defined by
 ${\cal Z}^{[R]}_{\xi}[J,\overline{\chi},\chi,\overline{\eta},\eta] $,
and changing the sources in (4.8) into
\begin{eqnarray*}
&&J'^{\nu}_b(x) = -\int d^4x'\, J_{a\mu}(x')\,{\rm i}\,
     [\overline{D}(x'\!-\!x)^{-1}]^{\mu \nu}_{ab})\,,
\\
&&\overline{\eta}'_{a\alpha}(x) = -\int d^4x'\,\overline{\eta}_{a'\alpha'}(x')
  \,{\rm i}\,[\overline{S}(x'\!-\!x)^{-1}]_{a'\alpha', a\alpha}\, ,
\\
&&\eta'_{a\alpha}(x) = -\int d^4x'\,{\rm i}\,
  [\overline{S}(x\!-\!x')^{-1}]_{a\alpha, a'\alpha'}\,\eta_{a'\alpha'}(x')\,,
\\
&&\overline{\chi}'_a(x) = -\int d^4x'\, \overline{\chi}_a'(x')
  \,{\rm i}\, [\overline{C}(x'\!-\!x)^{-1}]_{a' a}\, ,
\\
&&\chi'_a(x) = -\int d^4x'\,{\rm i}\,[\overline{C}(x\!-\!x')^{-1}]_{a a'}
   \,\eta_{a'}(x')\,,
\end{eqnarray*}
we can find, in the limit as $\xi \rightarrow 0 $, that
\begin{eqnarray}
&& S_{fi} = \langle f|
:\,{\rm exp} \Big\{\int\!d^4x \Big(
   - A_{a\mu}(x)\frac{\delta} {{\rm i} \delta J_{a\mu}(x)}
   + \overline{\psi}_{a\alpha}(x)
   \frac{\delta}{{\rm i}\delta \overline{\eta}_{a\alpha}(x)}
   - \psi_{a\alpha}(x) \frac{\delta}{(-{\rm i})\delta \eta_{a\alpha}(x)}
   \nonumber\\
&&\ \ \ \ \ \ +\overline{\omega}_a(x)
   \frac{\delta}{{\rm i}\delta \overline{\chi}_a(x)}
   - \omega_a(x) \frac{\delta}{(-{\rm i})\delta \chi_a(x)}
  \Big) \Big\} : |i \rangle\,
  X^{[R]}[J,\overline{\chi},\chi,\overline{\eta},\eta]
  \left. \right|_{J=\overline{\chi}=\chi=\overline{\eta}=\eta=0} \,, \ \ \
\end{eqnarray}
where $X^{[R]}[J,\overline{\chi},\chi,\overline{\eta},\eta]$
stands for the generating functional for truncated renormalized
Green functions $[10]$. This is the formular we need. In the above
manipulation, we have taken into account the fact that the momenta
of the particles in the initial or final states are on the mass
shell and for such momenta the renormalized full propagators
become the free propagators.
\par
\vspace{5mm}
\begin{center}
{\bf V}.\ \ Concluding Remarks
\end{center} \par
     Starting from the path integral of the generating functional for the
renormalized Green functions with the renormalized field functions
as basical variables and using the unperturbed Hamiltonian
operator that, under the Lorentz condition, describes the free
particles of the initial and final states in scattering processes,
we have formed a satisfying operator description and found the
total Hamiltonian which determine the time evolution of the
Heisenberg operators of the renormalized field functions. With the
help of the time evolution operator $U(\infty,-\infty)$ in the
interaction picture and of the Quasi-Gupta-Bleuler subspace formed
by the physical initial-final states, we have been able to easily
explain the definition and the unitarity of the scattering matrix
and found a formular to express the matrix elements in terms of
the truncated renormalized Green functions.
\par
    We have also briefly described the asymptotic field viewpoint based
on our operator description. It would be interesting to compare such a
viewpoint with that of the traditional asymptotic theory $[11-14]$.
\par \ \par
\vspace{4mm}
\begin{center}
\bf{ACKNOWLEDGMENTS}
\end{center} \par
    We are grateful to Professor Yang Li-ming for helpful discussions. This
work was supported in part by National Natural Science Foundation
of China (19875002) and  supported in part by Doctoral Program
Foundation of the Institution of Higher Education of China.
\par
\vspace{5mm}
\setcounter{equation}{0}
\def\theequation{A\arabic{equation}}
\begin{center}
{\bf Appendix} \ \ \ The Gupta-Bleuler Subspace of Initial-Final States
\end{center}
\par
    As was pointed out in section 2, the particles in the initial or final
states are described by the following Lagrangian and Lorentz condition:
\begin{eqnarray}
&& {\cal L}_0 = -\frac{1}{4} F^{[0]}_{a \mu \nu} F_a^{[0]\mu \nu}
  + \frac{1}{2} M^2 A_{a \mu}A_a^{\mu} + {\cal L}_{\psi}
  - \big( \partial_{\mu} \overline{\omega}_a(x) \big)
  \partial^{\mu} \omega_a(x) \,, \\
&& \partial^{\mu} A_{a \mu} = 0 \,,
\end{eqnarray}
According to the operator description in the text of this paper we should
keep the ghost term in equation (A1) and form the initial-final state
subspace by following the Gupta-Bleuler quantization method. To this end
we first use the Lorentz and modify (A1) into
\begin{eqnarray}
\widetilde{{\cal L}}_0(x) = {\cal L}_{GB}(x)
+ \frac{1}{2} M^2 A_{a \mu}(x) A_{a}^{\mu}(x) + {\cal L}_{\psi}(x)
- \big(\partial_{\mu}\overline{\omega}_a(x)\big) \partial^{\mu}\omega_a(x)\,,
\end{eqnarray}
where the Gupta-Bleuler Lagrangian ${\cal L}_{GB}(x)$ can be written as
$$
{\cal L}_{GB}(x)
       =  -\frac{1}{2}\big( \partial_{\mu} A_{a\nu}\big)
                      \big( \partial^{\mu} A_a^{\nu}\big) \,.
$$
Then we disregard the Lorentz condition and perform a canonical
quantization with $\widetilde{{\cal L}}_0(x)$ and define the
corresponding Hamiltonian operator $H_0$ and field operators.
Moreover we have to choose the so-called physical states and
define a physical initial-final state subspace with the help of
the Lorentz condition and the residual symmetry. We certainly
expect the ghost particles to be excluded from appearing in the
initial-final states.
\par
    Assume that the operators of $\omega_a(x)$ are anti-hermitian and
therefore the operators of $\overline{\omega}_a(x)$ is hermitian.
Let $\omega^{(1)}_a(x)$, $\omega^{(2)}_a(x)$ stand for
$\overline{\omega}_a(x)$, $i\omega_a(x)$ and $\Pi^{(1)}_a$,
$\Pi^{(2)}_a$, $\Pi_{a\mu}$ stand for the canonical momenta
conjugate to $\omega^{(1)}_a(x)$, $\omega^{(2)}_a(x)$, $A_a^{\mu}$
respectively. Thus the operators of these quantities in the
Schrodinger picture are
\begin{eqnarray}
&& A_{a\mu}({\bf x}) = \int \frac{d^3k}{\sqrt{2\Omega(2\pi)^3}}
        \Big[ a_{a\mu}({\bf k})e^{i{\bf k}\cdot {\bf x}}
        + a^{\dagger}_{a\mu}({\bf k})e^{-i{\bf k}\cdot {\bf x}} \Big]\,, \\
&& \Pi_{a\mu}({\bf x}) = \int \frac{id^3k}{\sqrt{(2\pi)^3}}
             \sqrt{\frac{\Omega}{2}}
        \Big[ a_{a\mu}({\bf k})e^{i{\bf k}\cdot {\bf x}}
        - a^{\dagger}_{a\mu}({\bf k})e^{-i{\bf k}\cdot {\bf x}} \Big]\,, \\
&& \omega^{(1)}_a({\bf x}) = \int \frac{d^3k}{\sqrt{2|{\bf k}|(2\pi)^3}}
    \Big[ \overline{C}_a({\bf k})e^{i{\bf k}\cdot {\bf x}}
    + \overline{C}_a^{\dagger}({\bf k})e^{-i{\bf k}\cdot {\bf x}} \Big]\,, \\
&& \Pi^{(1)}_a({\bf x}) = \int \frac{(-i)d^3k}{\sqrt{(2\pi)^3}}
             \sqrt{\frac{|{\bf k}|}{2}}
    \Big[ C_a({\bf k})e^{i{\bf k}\cdot {\bf x}}
    + C^{\dagger}_a({\bf k})e^{-i{\bf k}\cdot {\bf x}} \Big]\,, \\
&& \omega^{(2)}_a({\bf x}) = \int \frac{(-i)d^3k}{\sqrt{2|{\bf k}|(2\pi)^3}}
    \Big[ C_a({\bf k})e^{i{\bf k}\cdot {\bf x}}
    - C^{\dagger}_a({\bf k})e^{-i{\bf k}\cdot {\bf x}} \Big]\,, \\
&& \Pi^{(2)}_a({\bf x}) = \int \frac{d^3k}{\sqrt{(2\pi)^3}}
             \sqrt{\frac{|{\bf k}|}{2}}
    \Big[\overline{C}_a({\bf k})e^{i{\bf k}\cdot {\bf x}}
    -\overline{C}^{\dagger}_a({\bf k})e^{-i{\bf k}\cdot {\bf x}} \Big]\,, \\
&& H_0 = -g^{\mu\nu} \int d^3k \Omega(|{\bf k}|)
         a^{\dagger}_{a\mu}({\bf k}) a_{a\nu}({\bf k})
         + \int d^3k |{\bf k}| \Big\{
         \overline{C}^{\dagger}_a({\bf k}) C_a({\bf k})
         + C^{\dagger}_a({\bf k}) \overline{C}_a({\bf k}) \Big\}
         + H^0_{\psi}\,,
\end{eqnarray}
where
\begin{eqnarray}
&& \Omega(|{\bf k}|) = \sqrt{M^2+|{\bf k}|^2} \,, \\
&& \Big[a_{a\mu}({\bf k}), a^{\dagger}_{b\nu}({\bf k}')\Big]
    = - g_{\mu\nu} \delta_{ab}\delta^3({\bf k}-{\bf k}') \,, \\
&&  \Big[ a_{a\mu}({\bf k}), a_{b\nu}({\bf k}')\Big] =
\Big[a^{\dagger}_{a\mu}({\bf k}), a^{\dagger}_{b\nu}({\bf k}')\Big] = 0 \,,\\
&& \Big[\overline{C}_a({\bf k}), C^{\dagger}_b({\bf k}')\Big]_+
   = \Big[C_a({\bf k}), \overline{C}^{\dagger}_b({\bf k}')\Big]_+
   = \delta_{ab}\delta^3({\bf k}-{\bf k}') \,, \\
&& \Big[C_a({\bf k}), C_b({\bf k}')\Big]_+
  = \Big[\overline{C}_a({\bf k}), \overline{C}_b({\bf k}')\Big]_+
  = \Big[C_a({\bf k}), \overline{C}_b({\bf k}')\Big]_+ \nonumber\\
&&\ \ \ \ \ \ \ \ \ \ \ \ \ \ \ \ \ \ \ \ \ \ \
  = \Big[\overline{C}_a({\bf k}), \overline{C}^{\dagger}_b({\bf k}')\Big]_+
  = \Big[{C}_a({\bf k}), C^{\dagger}_b({\bf k}')\Big]_+ = 0 \,.
\end{eqnarray}
The restriction of the Lorentz condition on a physical state
$|\Psi_{ph}\rangle$ can be expressed as
\begin{eqnarray}
&& k^{\mu} a_{a\mu}({\bf k}) |\Psi_{ph}\rangle = 0 \ \ \ \ \ \
   (k^0=\Omega(|{\bf k}|))  \,.
\end{eqnarray}
For a single particle state with the polarization vector
$\varepsilon^{\mu}({\bf k})$
\begin{eqnarray}
&& \varepsilon^{\mu}({\bf k}, \lambda) a^{\dagger}_{a\mu}({\bf k}) |0 \rangle
\end{eqnarray}
Eq. (16) gives
\begin{eqnarray}
&& k_{\mu}\varepsilon^{\mu}({\bf k}, \lambda) = 0 \,.
\end{eqnarray}
According to this condition two transversal polarization states
($\lambda=1,2$) and a longitudinal polarization state
($\lambda=3$) are allowed to be present for each {\bf k}.
\par
    Since $\partial^{\mu}\partial_{\mu}\omega_a(x)=0$ the theory we are
treating is invariant under the infinitesimal transformation
\begin{eqnarray}
&&
\delta A_{a\mu}(x) = \delta \zeta \partial_{\mu} \omega_a(x) \,, \ \ \ \
\delta \omega_a(x) = 0\,,\ \ \ \  \delta \overline{\omega}_a(x) = 0\,,\ \ \ \
\delta \psi(x) = 0 \,,
\end{eqnarray}
where $\delta \zeta$ is an infinitesimal fermionic real constant. Similarly,
since $\partial^{\mu}\partial_{\mu}\omega_a(x) = 0$ the theory is also
invariant under the transformation
\begin{eqnarray}
&& \delta A_{a\mu}(x) = \delta \zeta \partial_{\mu} \overline{\omega}_a(x) \,,
   \ \ \ \
\delta \omega_a(x) = 0\,,\ \ \ \  \delta \overline{\omega}_a(x) = 0\,,\ \ \ \
\delta \psi(x) = 0 \,.
\end{eqnarray}
This is the residual symmetry mentioned above. Under such a kind of
transformations a physical state should become a equivalent state.
Particularly
$\varepsilon^{\mu}({\bf k},\lambda)\delta a^{\dagger}_{a\mu}({\bf k})
|0 \rangle$ should be equivalent to zero, where
$\delta a^{\dagger}_{a\mu}({\bf k})$ is determined by
\begin{eqnarray}
&& \delta A_{a\mu}({\bf x})
        = \delta \zeta \big[\partial_{\mu}\omega_a\big]_{t=0} \,,
\end{eqnarray}
or by
\begin{eqnarray}
&& \delta A_{a\mu}({\bf x})
      = \delta \zeta \big[\partial_{\mu}\overline{\omega}_a\big]_{t=0} \,.
\end{eqnarray}
From (A21), one has
\begin{eqnarray}
&& \delta a^{\dagger}_{a\mu}({\bf k})
= i \delta \zeta \widetilde{k}_{\mu} \sqrt{\Omega/|{\bf k}|}
    C^{\dagger}_a({\bf k})\,.
\end{eqnarray}
where the components of $\widetilde{k}_{\mu}$ are
\begin{eqnarray}
&& \widetilde{k}_0 = |{\bf k}|\,,\ \ \ \ \ \ \widetilde{k}_l = k_l \,.
\end{eqnarray}
Similarly from (A22), one has
\begin{eqnarray}
&& \delta a^{\dagger}_{a\mu}({\bf k})
= i \delta \zeta \widetilde{k}_{\mu} \sqrt{\Omega/|{\bf k}|}
    \overline{C}^{\dagger}_a({\bf k})\,.
\end{eqnarray}
It follows that  $C^{\dagger}_a({\bf k}) |0 \rangle $  and
$\overline{C}^{\dagger}_a({\bf k})|0\rangle$ should be equivalent to zero.
\par
   In conclusion, the initial-final state subspace is formed by all the
physical states. A physical state can only contain material particles,
transversal polarization or longitudinal polarization gauge Bosons and not
ghost particles.
\par
\newpage
\begin{center}
{\large \bf References}
\end{center}
\par  \noindent
[1]\ G.Curci and R.Ferrari, Nuovo Cim. {\bf 32}, 151(1976).
\par  \noindent
[2]\ I.Ojima, Z. Phys. {\bf C13},173(1982).
\par  \noindent
[3]\ A.Blasi and N.Maggiore, het-th/9511068; Mod. Phys. Lett.{\bf A11},
1665(1996).
\par  \noindent
[4]\ R.Delbourgo and G.Thompson, Phys. Rev. Lett. {\bf 57}, 2610(1986).
\par  \noindent
[5]\ M.Carena and C.Wagner, Phys. Rev. {\bf D37}, 560(1988).
\par  \noindent
[6]\ A.Burnel, Phys. Rev. {\bf D33}, 2981(1986);{\bf D33}, 2985(1986).
\par  \noindent
[7]\ T.Fukuda, M.Monoa, M.Takeda and K.Yokoyama, \par\ \ \ \
Prog. Theor. Phys. {\bf 66},1827(1981);{\bf 67},1206(1982);{\bf 70},284(1983).
\par  \noindent
[8]\ Yang Ze-sen, Zhou Zhining, Zhong Yushu and Li Xianhui,
     hep-th/9912046, 7 Dec 1999.
\par  \noindent
[9]\ Z.S.Yang, X.H.Li and X.L.Chen, hep-ph/0007007, 3 Jul 2000.
\par \noindent
[10]\ Z.N.Zhou, Y.S.Zhong and  X.H.Li, Chin.Phys.lett.${\bf 12}$,\,1(1995).
\par \noindent
[11]\ H.Lehmann, K.Symanzik and W.Zimmermann,\,Nuovo Cimento, ${\bf 1}$,\,205
(1955).
\par \noindent
[12]\ J.D.Bjorken and S.D.Drell,\,"Relativistic Quantum Field", McGraw-Hill,
 New York, 1965.
\par \noindent
[13]\ C.Itzykson and J.B.Zuber,\,"Quantum Field Theory", McGraw-Hill Inc.,
 New York, 1980.
\par \noindent
[14]\ Michio Kaku,\,"Quantum Field Theory", Oxford University Press,
 Oxford, 1993.
\par
\end{document}